\documentstyle[preprint,aps]{revtex}
\begin{document}
\title{Canonical Partition Functions for  Parastatistical  Systems
of any order}
\author{S. Chaturvedi}
\address{School of Physics\\
University of Hyderabad\\
Hyderabad - 500 046 (INDIA)}
\maketitle
\begin{abstract}
A general formula for the canonical partition function for a system obeying
any statistics based on the permutation group is derived. The formula expresses
the canonical partition function in terms of sums of Schur functions. The only
hitherto known result due to Suranyi [ Phys. Rev. Lett. {\bf 65}, 2329 (1990)]
for parasystems of order two is shown to arise as a special case of our
general formula. Our results also yield all the relevant information about the
structure of the Fock spaces for parasystems.
\end{abstract}
\vskip1.0cm
\noindent 19 September 1995
\newpage
\def\ch{{\cal H}}
There exist two approaches to parastatistics.  The  first  is  the
field theoretic approach$^{1-3}$ based on a generalization of  the
creation-annihilation operator algebra for  bosons  and  fermions.
This is the way in which parastatistics was  first  introduced  by
Green.$^1$ In  the  second  approach  pioneered  by  Messiah  and
Greenberg$^4$  and  further  investigated  by  Hartle  Stolt   and
Taylor,$^5$  parastatistics  arises  in  the  quantum   mechanical
description of an assembly of $N$-identical  particles.  Here  the
permutation group $S_N$ plays a central  role.  Of  the  two,  for
calculational purposes, the first  seems  to  have  found  greater
favour with the workers in this field.$^{6,7}$ Thus, for instance,
using this approach Suranyi$^7$ has given the  canonical  partition
function for a parabose and parafermi  gas  of  order  $p=2$.  The
calculation involves a clever use  of  the  simplifications  which
occur in the para algebra when the order of the parastatistics  is
two. In this letter we show that the quantum  mechanical  approach
to parastatistics when combined with the  machinery  of  symmetric
functions$^8$ yields  a  powerful  method  which  enables  one  to
answer,  with  great  facility,  all   questions   pertaining   to
statistics based on the permutation group. In particular, we  give
the canonical partition  function  for  a  parabose  or  parafermi
system of an arbitrary order.

To set the notation we begin with a brief summary of some familiar
results. This is also  necessitated  by  the  fact  that  it  is  an
in-depth appreciation and examination of what is all too  familiar
along with the intution gained in the process which leads  us,  in
one stroke to  the  desired  results.  Consider  a  Hilbert  space
${\ch}$ built by an $N$-fold tensor product  of  a  Hilbert  space
$H$  of  dim  $M$.  We   shall   assume   that   $M\ge   N$.   Let
$1,2,3,\cdots,M$ denote the basis vectors of $H$. The $M^N$  basis
vectors of $\ch$ correspond to each term in the product
\begin{equation}
{(1+2+\cdots+M)(1+2+\cdots+M)\cdots(1+2+\cdots   +M)\over    N~{\rm
factors}}\,\,\,\,.
\end{equation}
One may consider two decompositions of this set of $M^N$ states.
     \begin{enumerate}
          \item {\it Decomposition based on  occupation  numbers}:

	\end{enumerate}
		Here one groups together states  which  have  the  same
                number of 1's, 2's $\cdots$ etc., regardless of  their
                location  in  the  product.  Each  such  group  is
                characterized by a set of occupation numbers which
                give the number of times $1,2,\cdots  M$  occur  in
                the states in that group. All relevant aspects  of
                this  decomposition  are   encapsulated   in   the
                following decomposition of the symmetric  function
                $(x_1+\cdots + x_M)^N$,
                    \begin{equation}
                    Z_N^{inf}(x_1,\cdots,x_M)               \equiv
                    (x_1+x_2\cdots        +        x_M)^N        =
                    \sum_{\lambda\atop{|\lambda|=N}}
                    {N!\over\lambda_1!\cdots\lambda_M!}
                    ~m_\lambda(x_1,\cdots, x_M)\,\,\,\,.
                    \end{equation}
                Here    $\lambda\equiv(\lambda_1,\lambda_2,\cdots,
                \lambda_M)$,    $\lambda_1\ge\lambda_2\ge\lambda_3
                \cdots  \ge  \lambda_M$  is  a  partition  of  $N$
                (indicated       by       $|\lambda|=N$)        and
                $m_\lambda(x_1, \cdots, x_M)$  denotes  the  monomial
                symmetric  function$^8$   corresponding   to   the
                partition $\lambda$
                    \begin{equation}
                    m_\lambda(x_1,\cdots,x_M)        =        \sum
                    x_1^{\lambda_1}x_2^{\lambda_2}          \cdots
                    x_M^{\lambda_M}\,\,\,\,.
                    \end{equation}
                The sum on the R.H.S. of (3) is over all  distinct
                permutations   of    $(\lambda_1,\cdots,\lambda_M)$.
                Setting $x_1=x_2=x_M=1$ in (2) we obtain
                    \begin{equation}
                    M^N  =  \sum_\lambda  {N!\over\lambda_1~\cdots
                    \lambda_M!} ~m_\lambda (1,\cdots,1 )\,\,\,\,,
                    \end{equation}
                which    tells    us    that    each     partition
                $\lambda=(\lambda_1,\cdots,\lambda_M)$ corresponds
                to  $m_\lambda(1,\cdots,1)$  sets  of   occupation
                numbers  obtained  by  distinct  permutations   of
                $\lambda_i$'s   and   each   such   set   contains
                $N!/\lambda_1!\cdots\lambda_M!$ states. The number
                $m_\lambda (1,\cdots,1)$ is given by
                    \begin{equation}
                    m_\lambda(1,1,\cdots,1)       =       {M!\over
                    m_1!m_2!\cdots }\,\,\,\,,
                    \end{equation}
                where  $m_i$'s  denote   the   number   of   times
                $\lambda_i$'s  occur  in   the   given   partition
                $\lambda$.

                Note also that with the identification
                    \begin{equation}
                    x_1=e^{-\beta\epsilon_1}~~,~~x_2=e^{-\beta
                    \epsilon_2}, \cdots, x_M=e^{-\beta\epsilon_M}\,\,\,\,,
                    \end{equation}
                where $\epsilon_1,\epsilon_2\cdots,\epsilon_M $ are  taken  to
                denote   the    ``single    particle    energies''
                corresponding  to  the  states  $1,\cdots,  M$,  the
                symmetric  function   $Z_N^{inf}(x_1,\cdots,   x_M)$
                represents the partition function  of  the  system
                under  consideration.  This  fact  is  made   more
                explicitly by rewriting (2) as
                    \begin{equation}
                    Z_{N}^{inf}(x_1,\cdots,          x_M)          =
                    (x_1+\cdots+x_M)^N  =  \sum_{{n_i}\atop{\Sigma
                    n_i=N}}
                    ~{N!\over  n_1!n_2!\cdots   n_M!}   ~x_1^{n_1}
                    x_2^{n_2}\cdots x_M^{n_M}\,\,\,\,.
                    \end{equation}
                (The symbol $Z_N$ was introduced with  a  view  to
                emphasizing this point.  The  superscript  ``inf''
                indicates that we are dealing  with  the  infinite
                statistics$^9$)
\begin{enumerate}
                \item  {\it Decomposition  based  on  the   permutation
                group}:
     \end{enumerate}

In this decomposition we regard the $M^N$ states  as  the  carrier
space for an $M^N$ dimensional representation of  the  permutation
group $S_N$. This reducible representation can be decomposed  into
the irreducible representations of $S_N$ which, as is well  known,
are in one to one correspondence with the partitions  of  $N$.  In
this case, the relation analogous  to  (2)  which  summarises  all
relevant features of this decomposition is
\begin{equation}
Z_N^{inf}(x_1,\cdots,   x_M)   \equiv   (x_1+\cdots   +   x_M)^N   =
\sum_{{\lambda\atop |\lambda|=N}} n(\lambda)  S_\lambda  (x_1,\cdots,
x_M)\,\,\,\,,
\end{equation}
where $S_\lambda(x_1,\cdots,x_M)$ are the Schur functions$^8$
\begin{equation}
S_\lambda(x_1,\cdots,x_M) =  {\det(x_i^{\lambda_j+M-j)}\over  \det
(x_i^{M-j})} ~~ ; ~~ 1 \le i,j \le M \,\,\,\,,
\end{equation}
and  $n(\lambda)$  denotes  the  dimension  of   the   irreducible
representation $\lambda$ of $S_N$. Setting $x_1=x_2=\cdots=x_M=1$ in
(8) we get
\begin{equation}
M^N       =       \sum_{{\lambda\atop|\lambda|=N}}       n(\lambda)
S_\lambda(1,\cdots, 1)\,\,\,\,,
\end{equation}
which tells us that $S_\lambda(1,\cdots, 1)$ is the  number
of times the irreducible representation $\lambda$ occurs  in  this
decomposition.

So far we had been dealing with $\ch$. Following refs. 4 and 5  we
now construct out of it a generalized ray space $\ch_{phy}$ by
\begin{itemize}
\item{(a)} admitting only  those  operators  on  $\ch$  which  are
permutation symmetric,
\item{(b)} identifying those states in $\ch$ which have  the  same
expectation values for all permutation symmetric operators.
\end{itemize}

These assumptions, together with the Schur's  Lemma,  imply  that  all
states belonging to an  irreducible  representation  $\lambda$  of
$S_N$ count as one state of $\ch_{phy}$. This immediately tells  us
that  the  symmetric  function  of  degree  $N$   appropriate   to
$\ch_{phy}$ is simply obtained by setting $n(\lambda)=1$ in (8).
\begin{equation}
Z_N^{HST}  (x_1,\cdots,  x_M)  =  \sum_{{\lambda\atop|\lambda|=N}}
S_\lambda(x_1,\cdots, x_M)\,\,\,\,.
\end{equation}
This is the key result of this work. (Here we use the superscripts
HST  to  denote  Hartle  Stolt  and  Taylor  in  honour  of  their
contributions to parastatistics). This symmetric function captures
all   aspects   of   $\ch_{phy}$   much   the    same    way    as
$(x_1+\cdots+x_M)^N$ does for $\ch$.

So far no restrictions have been put on $\lambda$ --- the  sum  on
the R.H.S. of (11) is over all partitions of $N$. We  shall  refer
to this statistics as HST statistics. Parabose case of  order  $p$
arises when we retrict the sum in (11) to only  those  partitions
of  $N$  whose  length  $l(\lambda)$  (the  number   of   the   non-zero
$\lambda_i$'s) is less than  equal  to  $\le p$.  In  terms  of  Young
tableaux, this amounts to retaining only  those irreducible representations
pf $S_N$ in which the number of boxes in the first column is $\le  p$.  The
appropriate symmetric function for this case is
\begin{equation}
Z_N^{P.B}(x_1,\cdots, x_M  ;  p)  =  \sum_{\lambda{\atop |\lambda|=N}
\atop{l(\lambda)\le p}} S_\lambda (x_1,\cdots, x_M)\,\,\,\,.
\end{equation}
Similarly parafermi case of order  $p$  arises  when  we  restrict
$\lambda$ in (11) to those partitions  whose  conjugate  partition
$\lambda^\prime$  is  of  length  $\le  p$.  In  terms  of   Young
tableaux  this  implies   retaining   only   those   irreducible
representations of $S_N$ in which the number of boxes in the first  row  is
$\le p$. The symmetric function appropriate to this case is
\begin{equation}
Z_N^{PF}(x_1, \cdots, x_M  ;  p)  =  \sum_{{\lambda\atop |\lambda|=N}
\atop{l(\lambda^\prime)\le p}} S_\lambda(x_1,\cdots, x_M)\,\,\,\,.
\end{equation}
Likewise, for the $(p,q)$ statistics the corresponding  symmetric
function $Z_N^{(p,q)}(x_1,\cdots, x_M)$ is  obtained  by  restricting
the sum in (11) to those partitions for  which  $l(\lambda)\le  p$
and $l(\lambda^\prime) \le q$. It may be noted that if $p,q\ge N$,
all these cases reduce to HST.

The symmetric functions above contain all the relevant information
about the appropriate $\ch_{phy}$. For instance the  dimension  of
$\ch_{phys}$ is obtained by setting  $x_1=x_2=\cdots  x_M=1$.  The
appropriate canonical partition function is  obtained  by  setting
$x_1=e^{-\beta\epsilon_1}$, $x_2=e^{-\beta\epsilon_2}, \cdots, x_M
= e^{-\beta\epsilon_M}$. Some formulae which prove to be extremely
useful in carrying out the sums in (12) and (13) with restrictions
on the lengths of the partitions are as follows.$^8$
\begin{eqnarray}
S_\lambda(x_1,\cdots, x_M) & = &  \det  (h_{\lambda_i}-i+j)~~  ;  ~~
1\le i,j \le l(\lambda)\,\,\,\,,  \\
S_\lambda(x_1,\cdots, x_M) & = &  \det  (e_{\lambda^\prime_i}-i+j)~~  ;  ~~
1\le i,j \le l(\lambda^\prime)\,\,\,\,.
\end{eqnarray}

Here the complete symmetric functions $h_r(x_1,\cdots, x_M)$ and the
elementary symmetric functions $e_r(x_1,\cdots, x_M)$ are defined as
follows
\begin{eqnarray}
h_r(x_1,\cdots,  x_M)   &   =   &   \sum_{\lambda\atop{|\lambda|=r}}
m_\lambda (x_1,\cdots, x_M)\,\,\,\,, \\
e_r(x_1,\cdots,           x_M)           &            =            &
\sum_{i_1<i_2<\cdots<i_r} x_{i_1} x_{i_2}\cdots x_{i_r}\,\,\,\,.
\end{eqnarray}
Using these formulae one can  express  the  appropriate  symmetric
functions in terms of either $h$'s or $e$'s. As  an  illustration,
let us consider the Bose case. Here, since $l(\lambda)\le  1$,  we
have only  one  term  on  the  R.H.S.  of  (13)  corresponding  to
$\lambda=(N,0,0\cdots,0)$. Using (14) we obtain
\begin{equation}
Z_N^B(x_1,\cdots,x_M) = h_N(x_1,\cdots, x_M)\,\,\,\,.
\end{equation}
Similarly, for the Fermi case, one has
\begin{equation}
Z_N^F(x_1,\cdots, x_M) = e_N(x_1,\cdots, x_M)\,\,\,\,.
\end{equation}
Consider parabose of order 2. Using (12) and (14) we obtain
\begin{equation}
Z_N^{PB}(x_1,\cdots,         x_M;2)         =         h_N          +
\sum_{\lambda_1+\lambda_2=N\atop{\lambda_1\ge   \lambda_2}}   \det
\left(\begin{array}{ccc}
h_{\lambda_1} & h_{\lambda_1+1}\\
h_{\lambda_2-1} & h_{\lambda_2}\end{array}\right)\,\,\,\,,
\end{equation}
which on simplification leads to
\begin{eqnarray}
 & = & h^2_P(x_1,\cdots, x_M)~~~ {\rm if}
{}~~~ N=2P\,\,\,\,, \nonumber\\
Z_N^{PB}(x_1,\cdots, x_M ; 2)&& \\
& = & h_{P+1}(x_1,\cdots, x_M) h_P(x_1,\cdots, x_M) ~~~ {\rm  if}  ~~~
N=2P+1\,\,\,\,. \nonumber
\end{eqnarray}
The result for parafermi of order 2 is obtained by replacing $h$'s
by $e$'s. Thus we obtain the the  results  due  to  Suranyi  which
arise as a special case of (12) and (13). One can carry out similar
calculations for any order $p$.

For the HST case a number of interesting formulae can  be  derived
from the following result for the Schur functions$^8$
\begin{equation}
\sum_N \sum_{\lambda\atop{|\lambda|=N}}  S_\lambda(x_1,\cdots, x_M )  =
\prod_i{1\over(1-x_i)}\prod_{i<j} {1\over(1-x_i x_j)}\,\,\,\,.
\end{equation}
The R.H.S. of (22), with appropriate identifications of the $x_i$'s,
can be seen to be
the grand canonical partition function for the HST statistics. Further,
setting $x_1 = \cdots = x_M = t$ and reading off the coefficient of $t^N$
in the resulting expression on the R.H.S. one obtains
\begin{eqnarray}
\dim ({\cal  H}_{phys})  &  =  &  \sum_{\lambda\atop{|\lambda|=N}}
S_\lambda(1, \cdots,1)\nonumber\\
& = & \sum_{S=0}^{[N/2]}
\left( \begin{array}{cc}
M+N-2S-1\\
N-2S\end{array}\right)
\left(\begin{array}{cc}
M(M-1)/2+S-1\\
S\end{array}\right)\,\,\,\,.
\end{eqnarray}
This formula is interesting in its  own  right  as  it  gives  the
number  of  irreducible  representations of $S_N$ which   occur   in   the
decomposition of the $M^N$  dimensional reducible  representation  of  $S_N$
discussed above.

Finally, having derived  the  symmetric  function  appropriate  to
$\ch_{phy}$ (with or without restrictions on  the irreducible  representations
of the permutation  group)  we  can  immediately  obtain  all  the
relevant   information    regarding    its    occupation    number
decomposition. All  that  needs  to  be  done  is  to  expand  the
appropriate symmetric functions in terms of the monomial symmetric
functions     $m_\lambda(x_1,\cdots,      x_M)$.      Now      since
$S_\lambda(x_1,\cdots,   x_M)$   and   $m_\lambda(x_1\cdots   x_M)$;
$|\lambda|=N$,  serve  as  bases  for  symmetric  polynomials   in
$x_1,\cdots ,x_M$ of degree $N$, we can expand one in terms of  the
other
\begin{equation}
S_\chi(x_1,\cdots, x_M) = \sum_{\lambda\atop|\lambda|=N} K_{\chi\lambda}
m_\lambda(x_1,\cdots, x_M)\,\,\,\,,
\end{equation}
where    $K_{x\lambda}$    are    Kostka-Foulkes    numbers.$^{10}$
Substituting this in (11) we obtain
\begin{equation}
Z_N^{HST}(x_1,\cdots,  x_M)  =  \sum_\lambda  (\sum_\chi  K_{\chi\lambda})
m_\lambda (x_1,\cdots, x_M)\,\,\,\,.
\end{equation}
(For parabose  or  parafermi  or  $(p,q)$  case  one  has  to  put
appropriate  restrictions  on  the  partitions  $\chi$  in  (26)).  The
coefficients $\sum_\chi K_{\chi\lambda}$ immediately give us the  number
of   states    corresponding    to    the    occupation    numbers
$\lambda=(\lambda_1,\cdots, \lambda_M)$ or any distinct  permutation
thereof.

To conclude, by combining the approach propounded in refs.  4  and
5, with the theory of symmetric functions, we have  been  able  to
obtain  partition  functions  for  all  statistics  based  on  the
permutation  group.  Detailed  analyses   of   the   thermodynamic
properties  derivable  from  these  results  would  be   published
elsewhere.
\vskip0.35cm
\noindent{\bf Acknowledgements:} The author wishes to express  his
deep gratitude towards Prof.  V.  Srinivasan  for  his  continuous
encouragement and guidance. He  is  also   grateful  to  Prof.  C.
Musili, Dr. P.K. Panigrahi, Prof. A.K. Kapoor and  Prof.  R.
Jagannathan for valuable discussions and advice.
\newpage
\noindent{\bf References}
\begin{enumerate}
\item H.S. Green, Phys. Rev. {\bf90}, 270 (1953).
\item S.Doplicher, R. Haag and J.E.  Roberts,  Comm.  Math.  Phys.
      {\bf23}, 199 (1971).
\item Y. Ohnuki and S. Kamefuchi, {\it Quantum  field  theory  and
      parastatistics} (Springer Verlag, Berlin, 1982); S.N. Biswas
      in {\it Statistical Physics}, eds. N. Mukunda, A.K. Rajagopal  and
      K.P. Sinha, Proceedings of the symposium on fifty  years  of
      Bose statistics, I.I.Sc. Bangalore (India).
\item A.M.L. Messiah and O.W. Greenberg, Phys.  Rev.  B{\bf  136},
      248 (1964); O.W. Greenberg and A.M.L. Messiah,  B{\bf  138},
      1155 (1965).
\item J.B. Hartle and J.R.  Taylor,  Phys.  Rev.  {\bf178},  2043
      (1969); R.H. Stolt and J.R. Taylor, Phys. Rev. D{\bf1}, 2226
      (1970); J.B. Hartle, R.H. Stolt and J.R. Taylor, Phys.  Rev.
      D{\bf2}, 1759 (1970).
\item F. Mansouri and X. Wu, Phys. Lett.  B{\bf203},  417  (1988);
      S.N. Biswas and  A.  Das,  Mod.  Phys.  Lett.  A{\bf3},  549
      (1988); A. Bhattacharyya, F. Mansouri,  C.  Vaz  and  L.C.R.
      Wijewardhana, Phys. Lett. B{\bf224}, 384 (1989); Mod.  Phys.
      Lett. A{\bf12}, 1121 (1989).
\item P. Suranyi, Phys. Rev. Lett. {\bf65}, 2329 (1990).
\item I.G. Macdonald, {\it Symmetric  functions  and  Hall  polynomials}
      (Clarendon, Oxford, 1979).
\item O.W. Greenberg, Phys. Rev. Lett. {\bf64}, 705 (1990).
\item C. Kostka, Crell\'es  Journal  {\bf  93},  89  (1882);  H.O.
      Foulkes, {\it Permutations} (Gauthier-Villars, Paris, 1974).
\end{enumerate}
\end{document}